\documentclass[11pt,oneside]{amsart}

\usepackage[english]{babel}    
\usepackage[utf8]{inputenc}    
\usepackage[T1]{fontenc}       
\usepackage{amssymb} 		    
\usepackage{mathtools}         
\usepackage{icomma}            
\usepackage{units}             
\usepackage{enumerate}         
\usepackage{array}             
\usepackage[usenames]{color}   
\usepackage{graphicx}          

\usepackage{subcaption}

\usepackage{dsfont}
\usepackage{algorithm}
\usepackage{algpseudocode}

\usepackage[scale=.83]{geometry}

\usepackage[bookmarks, colorlinks=true, linkcolor=black, anchorcolor=black,
citecolor=black, filecolor=black, menucolor=black, runcolor=black,
urlcolor=black, pdfencoding=unicode]{hyperref}  

\theoremstyle{remark}

\newcommand{\RR}{\ensuremath{\mathbb{R}}}

\newcommand{\N}{\ensuremath{\mathbb{N}}}

\renewcommand{\P}{\ensuremath{\mathbb{P}}}

\newenvironment{itemize*}{\vspace{-10pt}\begin{itemize}\setlength{\itemsep}{0pt}\setlength{\parskip}{2pt}}{\end{itemize}}
\newenvironment{enumerate*}{\vspace{-10pt}\begin{enumerate}\setlength{\itemsep}{0pt}\setlength{\parskip}{2pt}}{\end{enumerate}}
\newenvironment{description*}{\vspace{-12pt}\begin{description}\setlength{\itemsep}{0pt}\setlength{\parskip}{2pt}}{\end{description}}

\newcommand{\Sp}{\ensuremath{\mathbb{S}}}
\newcommand{\abs}[1]{\left|#1\right|}
\newcommand{\set}[2]{\left\{#1\ ; \ #2  \right\}}
\renewcommand{\d}{\mathrm{d}}

\newcommand{\e}[1]{\vec{\mathrm{e}}_{#1}}

\title{An algorithm for simulating Brownian increments on a sphere}

\author{Aleksandar Mijatovi{\'c}}
\address{Department of Statistics, University of Warwick, \& The Alan Turing Institute, UK}
\email{a.mijatovic@warwick.ac.uk}
\author{Veno Mramor}
\address{Department of Statistics, University of Warwick, \& The Alan Turing Institute, UK}
\email{veno.mramor@warwick.ac.uk}
\author{Ger{\'o}nimo Uribe Bravo}
\address{Instituto de Matematicas, Universidad Nacional Aut{\'o}noma de M{\'e}xico, 
M{\'e}xico }
\email{geronimo@matem.unam.mx}

\numberwithin{equation}{section}
\numberwithin{thm}{section}

\begin{document}

\begin{abstract}
	This paper presents a novel formula for the transition density of the Brownian motion on a sphere of any dimension and discusses an algorithm for the simulation of the increments of the spherical Brownian motion based on this formula.
	The formula for the density is derived from an observation that a suitably transformed radial process (with respect to the geodesic
	distance) can be identified as a Wright-Fisher diffusion process. Such processes  satisfy a duality (a kind of symmetry) with a certain coalescent processes and this in turn yields a spectral representation of the
	transition density, which can be used for exact simulation of their increments using the results of Jenkins~and~Span\`{o}~(2017). The symmetry then yields the algorithm for the simulation of the increments of the
	Brownian motion on a sphere. We analyse the algorithm numerically and show that it remains stable when the time-step parameter is not too small.
\end{abstract}

\maketitle

\section{Introduction}

Brownian motion is essentially a continuous time symmetric random walk and is therefore of significant importance in science. Classically, Brownian motion is defined on a (flat) Euclidean space and this process is very well
understood. For many applications, however, it is better to model the state space as a curved surface or some other manifold and then  Brownian motion on the manifold is of interest. The definition of Brownian motion on any
Riemannian manifold is possible via the Laplace-Beltrami operator as in~\cite{hsumanifolds} and then a typical example is  Brownian motion on the sphere $\Sp^2(R)$ in the three dimensional Euclidean space $\RR^3$. The
spherical Brownian motion has been used to model fluorescent markers in cell membranes~\cite{KrishnaFluor}, motion of bacteria~\cite{BactLi1}, migration of marine animals~\cite{marinemigration} and to study the whole world
phylogeography~\cite{phylogeo}, to name a few examples.  Brownian motions on more general surfaces and manifold  are also of interest (cf.~\cite{fara}), but many manifolds (at least those with positive curvature) can be
locally approximated by a sphere, so we henceforth only focus on a spherical Brownian motion.

However, whereas the simulation of Euclidean Brownian increments 
is easy, as it reduces to the simulation of Gaussian random variables, the situation on a sphere is more
involved. There is no closed form for the transition density of a spherical Brownian motion, so one is usually forced to use
approximations. The first  approach is to locally approximate the sphere with a (flat) tangent space and then suitably project the Gaussian increment back to the sphere. This method was for example been applied
in~\cite{KrishnaFluor}, but its drawback is that it ignores the curvature of the sphere, so that it is a good approximation only for very small time-steps or very large radii. Later~\cite{sim4dimsphere}
provided a better approximation for the Brownian motion on the sphere in $\RR^4$ and later the algorithm has been adapted to work also on the sphere in $\RR^3$~\cite{sim2dimsphere}. 
In~\cite{gaussForSph} a different approximation is derived directly for the Brownian motion on $\Sp^2(R)$. These improved methods not only give good results for small time-steps, but continue to provide a good approximation
also for medium-sized time-steps. Nevertheless, all these algorithms are just approximate and as the time-step increases the results become less and less accurate. 

The main result in this article is a new representation of a transition density of the spherical Brownian motion given in Equation~\eqref{goodrho} and a numerical analysis of an algorithm for the simulation of the increments of spherical
Brownian motion arising from it. The representation is a consequence of the skew-product decomposition of the
spherical Brownian motion obtained in~\cite{simViaWF} and the representation of transition densities of Wright-Fisher diffusion processes obtained in~\cite{GriffLi83,exsimWF}. An algorithm for the simulation of the
increments of spherical Brownian motion is derived from the representation and is a generalization of~\cite[Algorithm~1]{simViaWF} allowing general diffusion coefficients and radii of the sphere. 

In contrast to the existing literature in which algorithms produce samples from an approximate distribution and
are hence forced to use smaller time-steps for accurate results, our algorithm is exact, so that it produces samples directly from the required distribution of the increments, and can handle
arbitrary large time-steps. In fact, its performance improves with increasing time-steps. The limiting factor of our algorithm is actually imperfect floating point arithmetic on computers. Since several quantities in the
algorithm have to be computed by complicated expression which  become numerically unstable for smaller time-steps, we are
actually limited with how small a time-step we are allowed to take. As such, our algorithm complements the other available methods. Moreover, in the numerical examples of Section~\ref{Comp_simulation} below it appears to
outperform them when the parameters are in the correct domain so we can use our algorithm. 

\section{Theory} We are interested in a Brownian motion on the sphere $$\Sp^{d-1}(R):=\set{\vec{x}\in\RR^d}{\abs{\vec{x}}=R},$$ where $\abs{\vec{x}}:=\sqrt{x_1^2+\cdots + x_d^2}$
is an absolute value of a vector $\vec{x}=(x_1,\ldots,x_d)^\top\in \RR^d$ so that $R>0$ represents the radius of the sphere
and $d\ge2$ is the dimension of the ambient Euclidean space. Hence, if we denote by
$\rho^{(D,R)}_{\vec{y}}(\vec{x},t)$ a transition density of the spherical Brownian motion started at $\vec{y}\in\Sp^{d-1}(R)$, we are interested in the diffusion equation:
\begin{equation} \label{diffeq}
\frac{\partial \rho^{(D,R)}_{\vec{y}}}{\partial t}(\vec{x},t)= D \nabla^2_{\Sp^{d-1}(R)}\rho^{(D,R)}_{\vec{y}}(\vec{x},t), \quad \rho^{(D,R)}_{\vec{y}}(\vec{x},0)=\delta(\vec{x},\vec{y}), 
\end{equation}
where $\nabla^2_{\Sp^{d-1}(R)}$ is the Laplace-Beltrami operator on the sphere $\Sp^{d-1}(R)$, $D>0$ is a
diffusion coefficient and $\vec{y}\in \Sp^{d-1}(R)$ is the initial point. Symmetry of  the spherical Brownian motion allows us to
deduce the following fact: if $A\in \RR^d\otimes\RR^d$ is any orthogonal matrix, then $\rho^{(D,R)}_{A\vec{y}}(A\vec{x},t)=\rho^{(D,R)}_{\vec{y}}(\vec{x},t).$ This in particular shows that it is enough to
consider a single initial point -- we are going to use the north pole $R\cdot\e{d}\in \Sp^{d-1}(R),$ where
$\e{d}:=(0,\ldots,0,1)^\top$ represents the north pole of the unit sphere. Furthermore, symmetry allows us to deduce that
$\rho^{(D,R)}_{\vec{y}}(\vec{x},t)$ does not depend on the whole vector $\vec{x}$ but only depends on the geodesic distance between the vectors $\vec{x}$ and $\vec{y}.$

We are using the standard round metric on the sphere which is induced by the standard Euclidean scalar product $\langle\vec{a},\vec{b}\rangle
=a_1b_1+\cdots+a_d b_d.$ The geodesic distance measures the shortest distance between two points (i.e. along
great circles which are intersections of the sphere with two dimensional planes through the origin) and is given as a function
$\operatorname{dis}\colon \Sp^{d-1}(R)\times \Sp^{d-1}(R)\to [0,R\pi], \operatorname{dis}(\vec{x},\vec{y})=R \arccos(\langle \vec{x},\vec{y}\rangle/R^2 )=:R\theta$ i.e. the distance is equal to $R$ times the angle between
the two vectors (here denoted by $\theta$). Usually, one looks at the standard spherical Brownian motion which diffuses on a
sphere $\Sp^{d-1}:=\Sp^{d-1}(1)$ of radius $1$ and corresponds to $D=1/2$ with initial point $\e{d}$. Since in
unit sphere coordinates relation  $\nabla^2_{\Sp^{d-1}(R)}=\nabla^2_{\Sp^{d-1}}/R^2$ holds, we can define a new parameter
$\tau:=2Dt/R^2,$ which we use instead of time and then $\rho^{(D,R)}_{R\e{d}}(\vec{x},t)=\rho^{(1/2,1)}_{\e{d}}(\vec{x}/ R,\tau)$ holds. This justifies the focus on the standard spherical Brownian motion since we can extend the results to
arbitrary radius and diffusion coefficient by rescaling and linear time-change. We will henceforth omit the
parameters $D=1/2$ and $R=1$ from the notation and simply write $\rho_{\e{d}}(\vec{z},\tau)$ for the transition density at time $\tau$ of  the standard spherical Brownian motion started at $\e{d}.$

Any point $\vec{z}\in \Sp^{d-1}$ can be represented as $\vec{z}=\sin{\theta}\vec{w}+\cos{\theta}\e{d},$ where
$\vec{w}\in\Sp^{d-2}\subseteq\RR^{d-1}\times\{0\}$ and $\theta\in[0,\pi]$ is the angle between the vectors $\vec{z}$ and $\e{d}$. Due to symmetry we see that
$\rho_{\e{d}}(\vec{z},\tau)=\rho_{\e{d}}(\sin{\theta}\vec{w}+\cos{\theta}\e{d},\tau)$ does not depend on the vector $\vec{w}$ but only depends on the angle $\theta$. Hence we can and will consider the one-dimensional density
\begin{align*}
\rho(\theta,\tau)&:=\mathrm{Vol}(\Sp^{d-2}(\sin\theta))\rho_{\e{d}}(\sin{\theta}\vec{w}+\cos{\theta}\e{d},\tau)=A_{d-2}\sin^{d-2}\theta\cdot\rho_{\e{d}}(\sin{\theta}\vec{w}+\cos{\theta}\e{d},\tau),
\end{align*}
where $A_n:=2\pi^{(n+1)/2}/\Gamma(\frac{n+1}{2})$ is the volume of an $n$-dimensional sphere $\Sp^n$.
The additional multiplicative factor $A_{d-2}\sin^{d-2}\theta$ is included due to only considering a
one-dimensional process (and its transition density) instead of the whole process on the sphere.
For $\tau=0$ the initial distribution $\rho(\theta,0)$ of this density is equal to the delta function with support at $0$. Using
(ultra)spherical harmonics and their addition formula we can get (see~\cite[p.~339]{KarTay}, where their Jacobi polynomials are normalised differently to our Gegenbauer polynomials) a formal solution of the diffusion equation at time $\tau$:
\begin{equation} \label{viaGeg}
\rho(\theta,\tau)= A_{d-2}\sin^{d-2}\theta\sum_{l=0}^\infty \frac{h(l,d)}{A_{d-1} }\frac{C^{(d/2-1)}_l(\cos\theta)}{C^{(d/2-1)}_l(1)} e^{\lambda_l\tau/2},
\end{equation}
where $C_n^{(\alpha)}$ are Gegenbauer polynomials given by their generating series $(1-2xt+t^2)^{-\alpha}=\sum_{n=0}^\infty C_n^{(\alpha)}(x)t^n$ and $h(l,d)=\binom{d+l-1}{d-1}-\binom{d+l-3}{d-1}$ is the dimension of the space of spherical harmonics of degree $l$, corresponding to the eigenvalue $\lambda_l:=-l(l+d-2).$
Equation \eqref{viaGeg} without the
additional factor $A_{d-2}\sin^{d-2}\theta$ (i.e. the density for the process on the sphere) has also been derived in~\cite{Cail2004} where a further simplification $h(l,d)/C^{(d/2-1)}_l(1)=(2l+d-2)/(d-2)$ is used.

In the special case $d=3$, i.e. for the standard unit sphere $\Sp^2$, equation~\eqref{viaGeg} specializes to the well-known formula
\begin{equation} \label{viaLeg}
\rho(\theta,\tau)=\frac{\sin\theta}{2}\sum_{l=0}^\infty(2l+1)P_l(\cos\theta)e^{-l(l+1)\tau/2},
\end{equation}
where $P_l$ is the $l$-th Legendre polynomial. Unfortunately, formal solutions given by formulas \eqref{viaGeg} or \eqref{viaLeg} oscillate a lot even when large amount of terms are taken into account and are therefore not suitable for
simulation. One is then usually forced to use some approximative density and sampling from it.

The simplest option is to locally approximate a sphere with its tangent space and use a standard Brownian increment which is then translated to an
increment on a sphere using exponential map i.e. moving an appropriate distance along the geodesics (great circles) of a sphere. Such an approximation corresponds to density
$Q_{Tang}(\theta,\tau)=\mathcal{N}_1(\tau)\theta^{d-2} \exp(-\theta^2/(2\tau)),$ where $\mathcal{N}_1(\tau)$ is a normalization constant such that $\int_0^\pi Q_{Tang}(\theta,\tau)\d\theta=1.$
This approximation largely ignores the curvature of the sphere so it is a good
approximation only for very small values of parameter $\tau$. Therefore, \cite{gaussForSph}  proposed an improved approximation. It is given by $Q_{Approx}(\theta,\tau)=\mathcal{N}_2(\tau)(\theta\sin\theta)^{(d-2)/2}
\exp(-\theta^2/(2\tau)),$ where $\mathcal{N}_2(\tau)$ is a normalization constant. This new approximation continues to
give good results even for an intermediate values of $\tau.$  In the case of $d=4$ the same approximative density was derived in~\cite{sim4dimsphere} by using a different method. 

The main result in this paper is the following alternative representation of a solution of the diffusion equation, which is much more suitable for simulation. It also allows us to do exact simulation in which we are sampling directly from
the transition density and not just from an approximative density as in the other previously mentioned methods. The alternative representation of the density~\eqref{viaGeg} is given by
\begin{equation}
\label{goodrho}
\rho(\theta,\tau)=\frac{\sin\theta}{2}\sum_{m=0}^{\infty}q^{d-1}_m(\tau)g_{(\frac{d-1}{2},\frac{d-1}{2}+m)}\left(\frac{1-\cos\theta}{2}\right), 
\end{equation}
where 
\begin{equation*}
g_{(\alpha,\beta)}(x):=\frac{1}{\operatorname{B}(\alpha,\beta)}x^{\alpha-1}(1-x)^{\beta-1}, \quad  x \in [0,1], \alpha,\beta>0
\end{equation*}
in which $\operatorname{B}$ represents the Beta function so that $g_{(\alpha,\beta)}$ is a density of a Beta distribution and
\begin{align}
\label{coef:q}
q^{d-1}_m(\tau)&:=\sum_{k=m}^\infty(-1)^{k-m}b^{(\tau,d-1)}_k(m),\\
b^{(\tau,d-1)}_k(m)&:=\frac{(d+2k-2)(d-1+m)_{(k-1)}}{m!(k-m)!}e^{-k(k+d-2)\tau/2}, \nonumber
\end{align}
where the Pochhammer symbol is given by $a_{(x)}=\frac{\Gamma(a+x)}{\Gamma(a)}$ for $a>0, x\ge-1.$

As a first impression the representation~\eqref{goodrho} looks just as unwieldy, if not  more,  than Equation~\eqref{viaGeg}, but
it does have a nice probabilistic interpretation, which makes it suitable for simulation. First,  the
coefficients $ q^{d-1}_m(\tau)$  are actually non-negative and they sum up to 1 (see  Subsection~\ref{derivation} below). Additionally, each term in the sum corresponds to  the density of a $\mathrm{Beta}(\frac{d-1}{2},\frac{d-1}{2}+m)$
distribution after we have used a transformation of the form $h(x)=\arccos(1-2x)$ (the inverse transformation is $h^{-1}(\theta)=\frac{1-\cos\theta}{2}$ and its derivative is exactly the factor $\frac{\sin\theta}{2}$ in front
of the sum). Therefore, we see that $\rho(\theta,\tau)$ is a mixture of densities of transformed Beta distributions.

This
immediately yields an algorithm for the simulation from the density $\rho(\theta,\tau).$ If we denote by $A_\infty^{d-1}(\tau)$ an $\mathbb{N}_0$-valued random variable with
$\mathbb{P}[A_\infty^{d-1}(\tau)=m]=q^{d-1}_m(\tau)$, then we can use the standard inversion sampling to simulate $A_\infty^{d-1}(\tau)$ i.e. for a random variable $U\sim \mathrm{Uniform(0,1)}$, the random variable 
$\inf \left\{ M\in\N_0; \sum_{m=0}^M q^{d-1}_m(t)>U \right\}$ is distributed as $A^{d-1}_\infty(t).$ Given this
value, we then simulate an appropriate Beta distributed random variable and after a final transformation we have found our sample from the density $\rho(\theta,\tau)$. 

Our goal is actually to simulate an $\Sp^{d-1}(R)$ valued random variable
correctly distributed as the increment of the Brownian motion on the sphere for an arbitrary initial point, radius and diffusion coefficient. Therefore, we have to recall symmetries of spherical Brownian motion i.e. $\rho^{(D,R)}_{A\vec{y}}(A\vec{x},t)=\rho^{(D,R)}_{\vec{y}}(\vec{x},t)$ for any orthogonal matrix $A$. In particular, 
by taking an orthogonal matrix which fixes the initial point, it implies that there is a component to the increment, which is uniformly distributed on $\Sp^{d-2}$.  Combining this fact with the equality
$\rho^{(D,R)}_{R\e{d}}(\vec{x},t)=\rho_{\e{d}}(\vec{x}/ R,\tau)$, where $\tau=2Dt/R^2$, and the aforementioned
simulation from density $\rho(\theta,\tau)$ we get the following algorithm for the exact simulation of the increment of the spherical Brownian motion. 

\begin{algorithm}[H]\caption{Simulating from the transition density  $\rho^{(D,R)}_{\vec{y}}(\, \cdot \,;t)$ of spherical Brownian motion} \label{simSpB}
	\begin{algorithmic}[1]
		\State Set $R=\abs{\vec{y}},\tau=2Dt/R^2$
		\State Simulate $M\sim A^{d-1}_\infty(\tau)$ \label{stepAinf}
		\State Simulate $X\sim \mathrm{Beta}(\frac{d-1}{2},\frac{d-1}{2}+M)$ \label{stepBeta}
		\State \label{step:uniform_sphere}Simulate $Y$ uniformly distributed on $\Sp^{d-2}$ 
		\State Set $O(\vec{y})=I-2uu^\top,$ where $u=(\e{d}-\vec{y}/R)/\abs{\e{d}-\vec{y}/R}.$ \label{stepOrt}
		\State \textbf{return} $R O(\vec{y})(2\sqrt{X(1-X)}Y^\top,1-2X)^\top$
	\end{algorithmic}
\end{algorithm}
Step~\ref{step:uniform_sphere} in Algorithm~\ref{simSpB} consists of simulating a vector $N$ in $\RR^{d-1}$
with independent standard normal components and setting $Y=N/\abs{N}$. 
The key property of the orthogonal matrix $O(\vec{y})\in \RR^d\otimes\RR^d$ in Algorithm~\ref{simSpB}
is $O(\vec{y})\e{d}=y/\abs{y}$. 
In fact, any orthogonal matrix in $\RR^d\otimes\RR^d$ with this property would lead to an exact sample.
The formula for $O(\vec{y})$ in Algorithm~\ref{simSpB} is chosen due to
its simplicity. Additionally, if only the simulation from the density $\rho(\theta,\tau)$ is required, we only
have to do first three steps and then random variable $\arccos(1-2X)$  has the adequate law. Step~\ref{stepAinf} constitutes the
main technical difficulty of the Algorithm~\ref{simSpB} and all potential numerical issues stem from it. By taking $R=1$ and $D=1/2$ we see that this algorithm is a more general version of~\cite[Algorithm~1]{simViaWF}.

\subsection{Derivation of the alternative representation of the density in \eqref{goodrho} and discussion} \label{derivation}
We will prove that the density  $\rho(\theta,t)$ is indeed represented as in \eqref{goodrho} and hence  justify the correctness of Algorithm~\ref{simSpB}. To do that we have to notice that density $\rho(\theta,t)$ is actually  the
transition density of  the  one-dimensional $[0,\pi]$-valued process $Y_t=\operatorname{dis}(Z_t,\e{d}),$ where $Z_t$ denotes the standard Brownian motion on the unit sphere $\Sp^{d-1}$ started at $\e{d}$. Since $\cos(Y_t)$
is nothing but the last component of the spherical Brownian motion $Z_t$, it is quite natural to further transform the process $Y_t$
and instead look at $X_t=\frac{1-\cos(Y_t)}{2}.$ This process is $[0,1]$-valued and surprisingly turns out to be
a Wright-Fisher diffusion process, a member of a family of processes which are well studied and used extensively in genetics. Computations in~\cite{simViaWF} show
that in our case the process $X_t$ is a Wright-Fisher diffusion process with parameters
$\theta_1=\frac{d-1}{2}=\theta_2$ and an initial point $y=0$ (see also~\cite[Eqs.~(1),(3)]{exsimWF}). Its transition density $f(x;t)$ is a solution of a Fokker-Planck equation 
\begin{align*}
\frac{\partial f(x;t)}{\partial t}= \frac{\partial^2}{\partial x^2}\left( \frac{x(1-x)}{2}f(x;t)\right)-\frac{\partial }{\partial x}\left(\left(\frac{d-1}{4}(1-x)-\frac{d-1}{4}x\right)f(x;t)\right), \quad f(x;0)=\delta(x,0).
\end{align*}

The transition densities of the Wright-Fisher diffusion processes have a spectral decomposition given by the associated Jacobi polynomials~\cite{GriffSpanoDiffCoal}, but such a representation is not suitable for simulation
(it is essentially equivalent to representation~\eqref{viaGeg}). Fortunately, there exists another representation of their transition
densities, which arises from the moment duality between Wright-Fisher diffusion processes and certain coalescent processes. This representation and its use for simulation was given in~\cite{GriffLi83} and it is given 
explicitly in~\cite[Eq.~(4)]{exsimWF}. The transformation formula for densities immediately yields
$\rho(\theta,\tau)=\frac{\sin\theta}{2}f\left(\frac{1-\cos\theta}{2};\tau\right)$ and combining it with~\cite[Eq.~(4)]{exsimWF} we immediately get the representation~\eqref{goodrho}.

Returning to the Algorithm~\ref{simSpB} we see that the steps~\ref{stepAinf} and~\ref{stepBeta} represent just a simulation from the density $f\left(\, \cdot\, ; \tau\right)$ of a Wright-Fisher diffusion process and are just
a particular case of~\cite[Algorithm~1]{exsimWF}. Furthermore, there is a more probabilistic description of the coefficients $q_m^{d-1}(\tau)$. 
Let $n\ge0$ and $\{A^{d-1}_n(\tau);\tau\ge0\}$ be a pure
death process on the non-negative integers, started at $A^{d-1}_n(0)=n$, where the only transitions are of the form
$m\mapsto m-1$ and occur at a rate $m(m+d-2)/2$ for each $m\in\{1,\ldots, n\}$. 
Then the coefficients $q^{d-1}_m(\tau)$ can be expressed as the limit 
$q^{d-1}_m(\tau)=\mathbb{P}[A_\infty^{d-1}(\tau)=m]=\lim_{n\to \infty}\P[A^{d-1}_n(\tau)=m]$.  This in particular shows that coefficients $q^{d-1}_m(\tau)$ are indeed non-negative and sum up to $1$.

The coefficients $q^{d-1}_m(t)$ are given only as an infinite series as in~\eqref{coef:q} and not as a closed expression, so the exact simulation of $A^{d-1}_\infty(\tau)$ is not trivial, but it is possible and is achieved
in~\cite{exsimWF}. For our purposes, sufficient accuracy is achieved simply by pre-computing the coefficients up to a certain precision and then using standard inversion sampling. 
\section{Computer simulations} \label{Comp_simulation}
\begin{figure}[!ht]
	\begin{subfigure}{.5\textwidth}
		\centering
		\includegraphics[width=.8\linewidth]{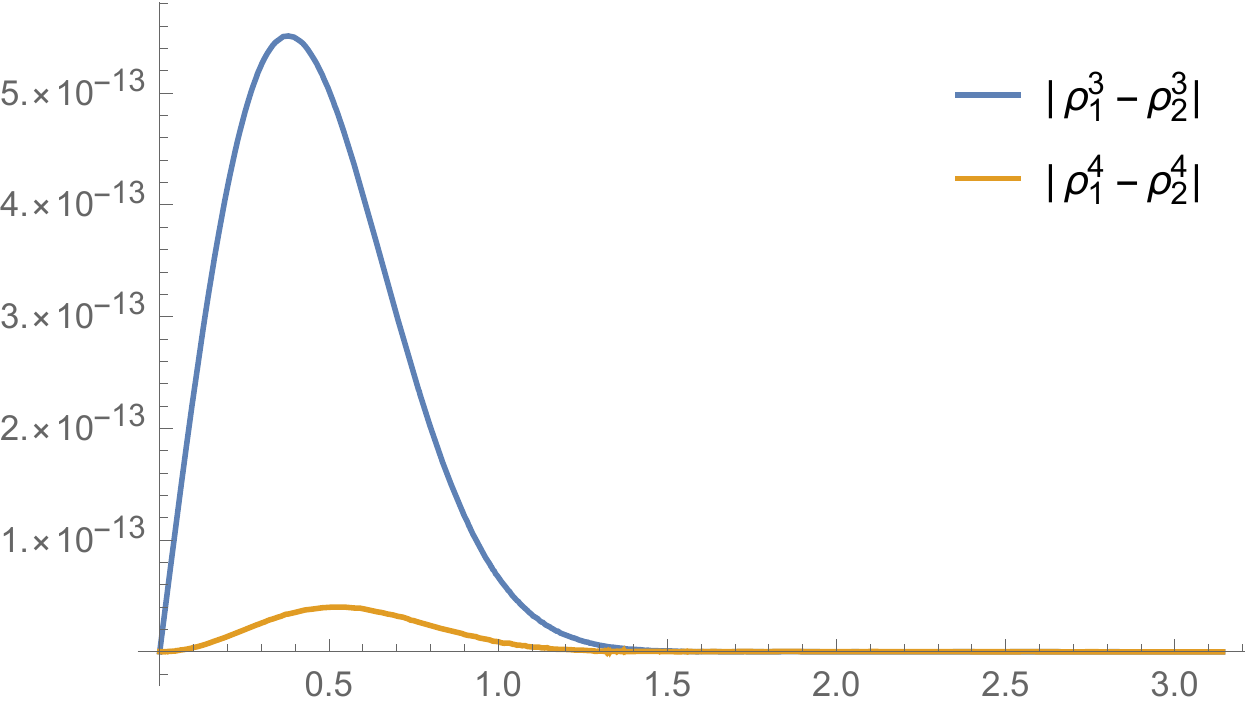}
	\end{subfigure}%
	\begin{subfigure}{.5\textwidth}
		\centering
		\includegraphics[width=.8\linewidth]{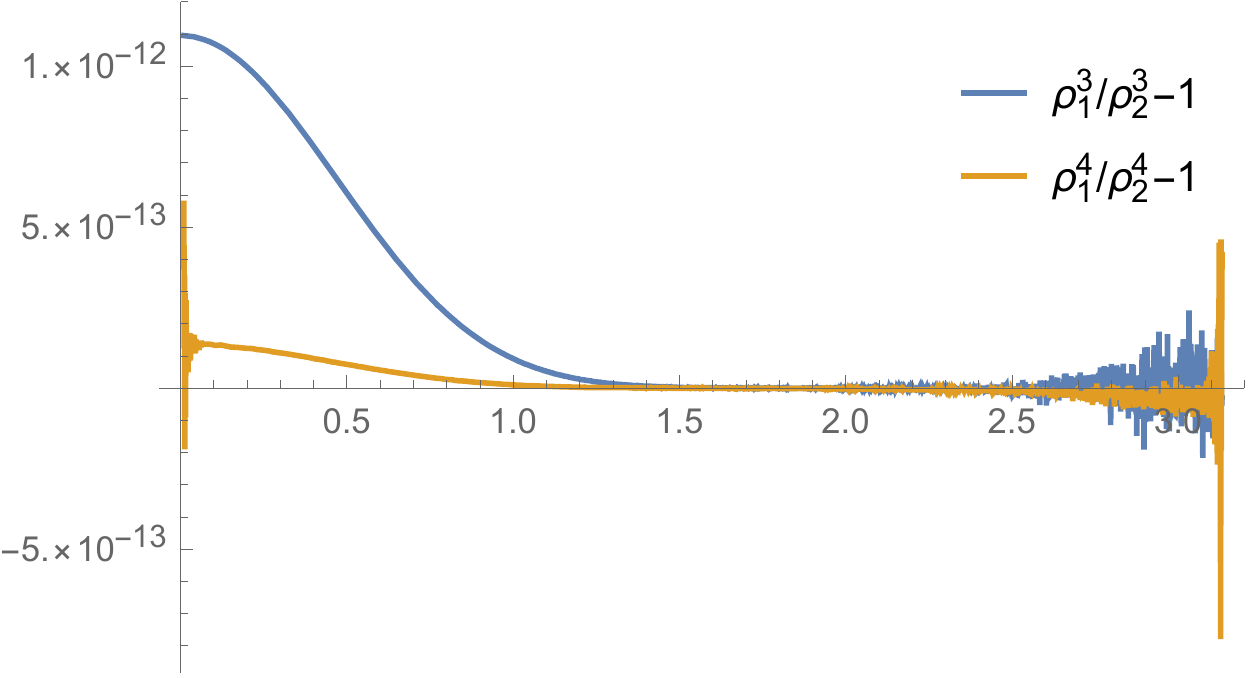}
	\end{subfigure}
	\caption{Absolute errors $\abs{\rho^d_1-\rho^d_2}$ on the left plot and relative error $\rho^d_1/\rho_2^d-1$ on the right plot for $d=3,4$.}
	\label{fig:errors}
\end{figure}

First, we check numerically that our alternative representation \eqref{goodrho} is actually equivalent to a more well-known representation \eqref{viaGeg}. To see that we are going to let $\tau=0.5$  and denote by $\rho_1^d$ the
density computed by the first $60$ terms in \eqref{viaGeg} and by $\rho_2^d$ density computed by the first $40$ terms in \eqref{goodrho} where each coefficient $q^{d-1}_m(\tau)$ is calculated using the first $60$ terms in \eqref{coef:q}. All
the densities have been calculated in Mathematica for $d=3,4$ and on Figure~\ref{fig:errors} we plot the absolute and relative error. Both errors are extremely small across the interval $[0,\pi]$, which confirms that both
densities are equal. Additionally, increasing the dimension $d$ seems to improve results.

Typically, when using algorithms for the simulation of the spherical Brownian motion, we need to repeatedly get samples for a fixed dimension $d$ and time $\tau.$ Therefore, first step is to pre-compute coefficients
$q_m^{d-1}(\tau)$ to a prescribed precision, so that we can afterwards use them in an inversion sampling of $A_\infty^{d-1}(\tau).$
Since the algorithm depends deeply on the representation \eqref{viaGeg}, it suffers from a numerical instability as $\tau\to 0.$ In theory the algorithm works correctly for all values of $\tau$, but in practice we are limited
by running time and by imperfect floating number precision, and whence the potential incorrect computations of the relevant
coefficients $q^{d-1}_m(\tau)$. When $\tau$ is small we have to compute products of the form $\frac{(d+2k-2)(d-1+m)_{(k-1)}}{m!(k-m)!}e^{-k(k+d-2)\tau/2}$ for large values of $k$ and $m$, which means
multiplying very large and very small numbers. This quickly accumulates numerical errors which cause the algorithm to fail. For
example, using direct calculations done in Mathematica via partial sums and the equation~\eqref{coef:q} we
calculate $q_{26}^{2}(0.05)$ to be equal to $-0.00427$ and even more extremely $q_{26}^{2}(0.03)$ to be equal to $29741.98$. Both of these results are clearly wrong and will cause the algorithm to fail.

In order to understand better how computation of coefficients $q_m^{d-1}(\tau)$ can go wrong, we have to look at the terms $b^{(\tau,d-1)}_k(m)$ from equation~\eqref{coef:q}. We need to be able to quantify the error in our
calculations, meaning we need to know how well the partial  sums $\sum_{k=m}^{m+K}(-1)^{k-m}b^{(\tau,d-1)}_k(m)$ approximate the
coefficients $q_m^{d-1}(\tau)$.   The partial sums form an alternating series and one is tempted to say that the
partial sums with even $K$ are always an upper bound for $q_m^{d-1}(\tau)$, whereas the partial sums with odd $K$ are always a
lower bound. Unfortunately, for this to hold, we would need  $b^{(\tau,d-1)}_k(m)\downarrow 0$ as $k\to \infty,$ which is not the case, although it is true once $k$ is large enough. To see this we notice that
$$\frac{b^{(\tau,d-1)}_{k+1}(m)}{b^{(\tau,d-1)}_k(m)}=\frac{m+k+d-2}{k-m+1}\frac{2k+d}{2k+d-2}e^{-(2k+d-1)\tau/2},$$
which is a product of three decreasing functions in $k$ and the limit as $k\to \infty$ is equal to $0$ (recall
that only terms with $k\ge m$ are relevant). This means that as soon as the quotient is for the first smaller than $1$ then the
terms $b^{(\tau,d-1)}_k(m)$ start decaying indefinitely. Therefore, for a large enough number of terms taken, the
exact value $q_m^{d-1}(\tau)$ always lies between two consecutive partial sums and they differ by just $b^{(\tau,d-1)}_k(m)$,
which is hence also the upper bound for the error.  This means that in order to compute coefficients $q_m^{d-1}(\tau)$ to a fixed precision $\varepsilon>0$ we simple start computing each coefficient via partial
sums in an infinite series representation~\eqref{coef:q} and we stop once the terms $b^{(\tau,d-1)}_k(m)$ start decaying and additionally $b^{(\tau,d-1)}_k(m)<\varepsilon$ holds.

\begin{figure}[!ht]
	\begin{subfigure}{.5\textwidth}
		\centering
		\includegraphics[width=.8\linewidth]{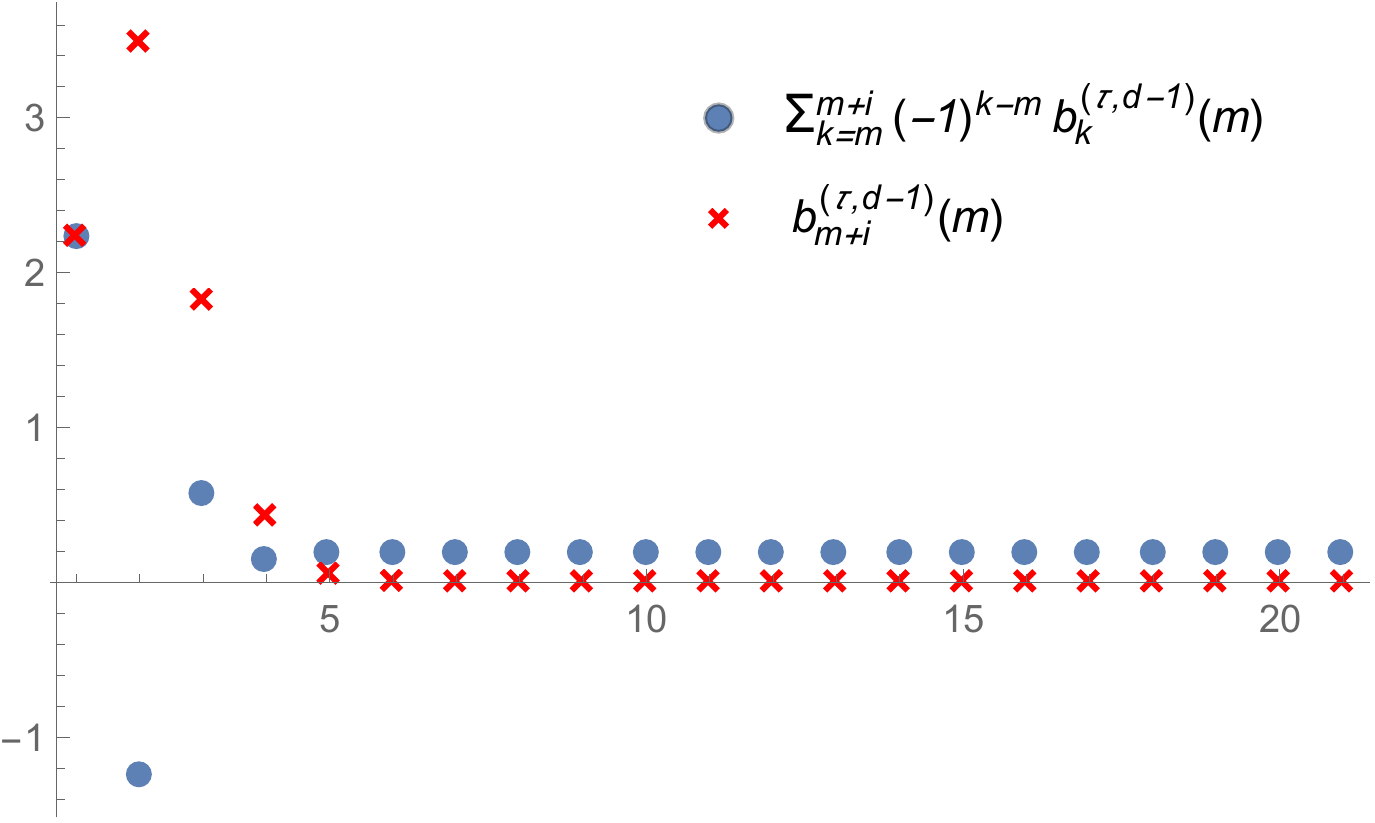}
	\end{subfigure}%
	\begin{subfigure}{.5\textwidth}
		\centering
		\includegraphics[width=.8\linewidth]{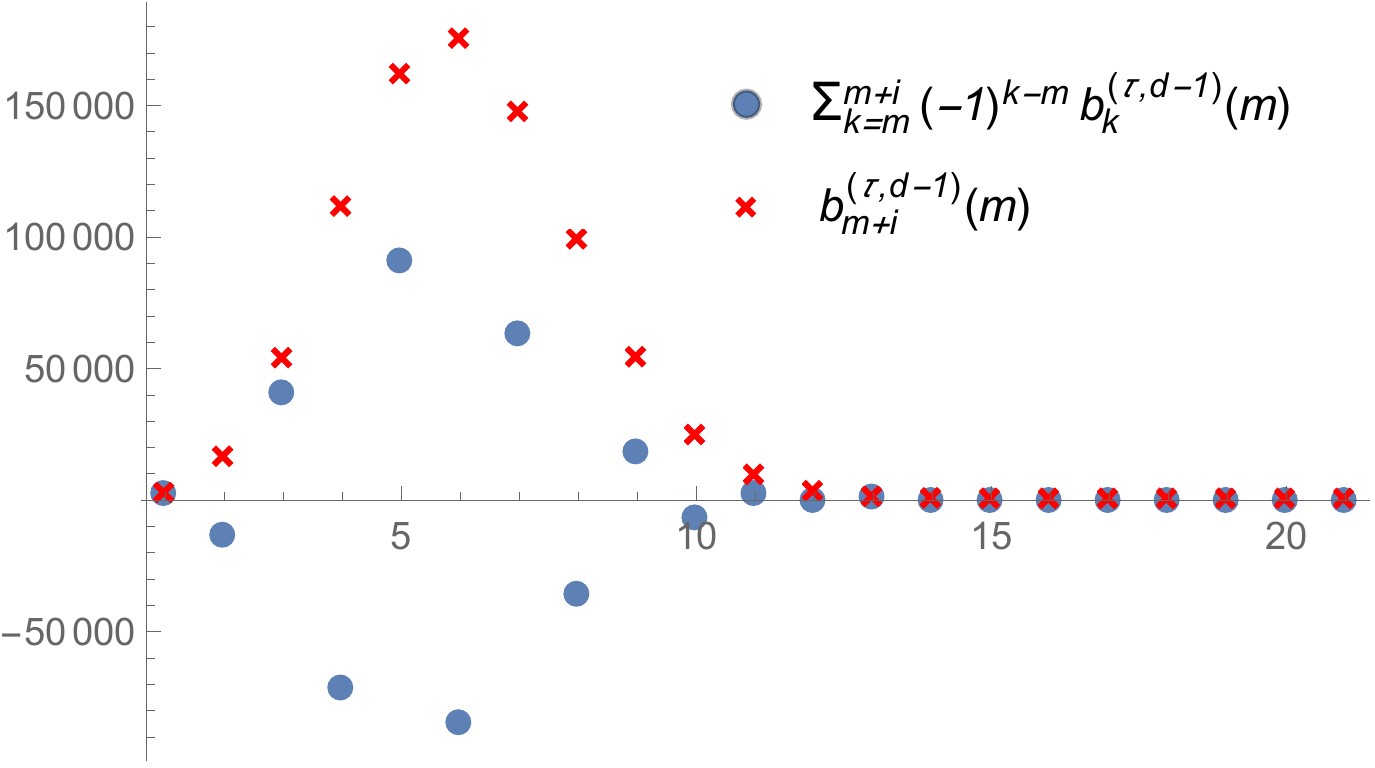}
	\end{subfigure}
	\caption{Plots of values $b^{(\tau,d-1)}_{m+i}(m)$ and  of partial sums $\sum_{k=m}^{m+i}(-1)^{k-m}b^{(\tau,d-1)}_k(m)$ for $i=0,\ldots,20$ and parameters for the left plot: $d=3,\tau=0.5,m=2$ and for the right plot: $d=3,\tau=0.1,m=13$.}
	\label{fig:coefbandpart}
\end{figure}

As we have seen above, the partial sums $\sum_{k=m}^{m+K}(-1)^{k-m}b^{(\tau,d-1)}_k(m)$ eventually become
arbitrarily close to the precise value $q_m^{d-1}(\tau)\in [0,1]$. However, before reaching this value they can sometimes be very
far off, in some cases the difference can be of several orders of magnitude. To see this we plot in
Figure~\ref{fig:coefbandpart} some of the terms $b^{(\tau,d-1)}_k(m)$ and some partial sums, which are used in the computation of $q_m^{d-1}(\tau)$.

One of the things immediately recognisable on Figure~\ref{fig:coefbandpart} is unimodality of terms $b^{(\tau,d-1)}_{k}(m)$, which is obvious from calculations in the previous paragraph. Another noticeable thing
is the size of the terms. Whereas on the left plot, where time $\tau$ is not too small, values of terms $b^{(\tau,d-1)}_k(m)$
and partial sums are relatively small, we can see that on the right plot, where the time $\tau$ is smaller,
values become really large, in our particular case they are of order $10^5$. We have to contrast this to the fact, that for the
values of parameters in the right plot the corresponding coefficient $q_{13}^{2}(0.1)$ is approximately equal to $0.00688$ which is of several orders of magnitudes smaller than the terms (and partial sums) used to calculate it. When the difference between the size of terms and the correct value becomes too large, numerical inaccuracies quickly add up
and numerically computed coefficients $q_m^{d-1}(\tau)$ are far from the correct values. This problem becomes even more apparent as the time $\tau$ goes to zero. Inspecting coefficients $q_m^{d-1}(\tau)$ suggests that the smallest time for
which calculations of coefficients still seem to go through without major numerical errors is around $\tau=0.1$ and the dimension $d$ doesn't seem to affect this as much (increasing dimension actually extends possible times $\tau$ for which the algorithm works correctly).  

\begin{figure}[!ht]
	\begin{subfigure}{.5\textwidth}
		\centering
		\includegraphics[width=.8\linewidth]{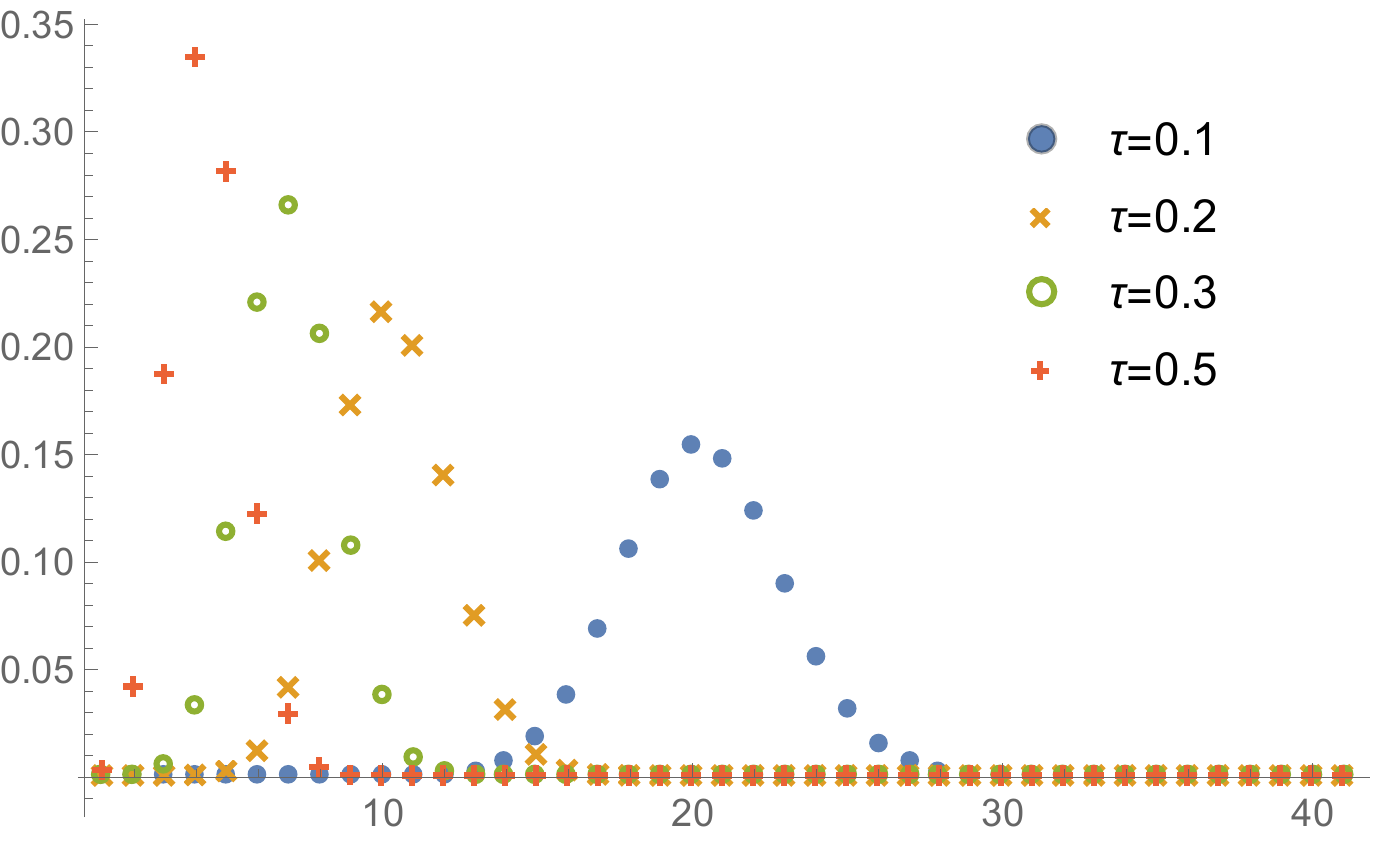}
	\end{subfigure}%
	\begin{subfigure}{.5\textwidth}
		\centering
		\includegraphics[width=.8\linewidth]{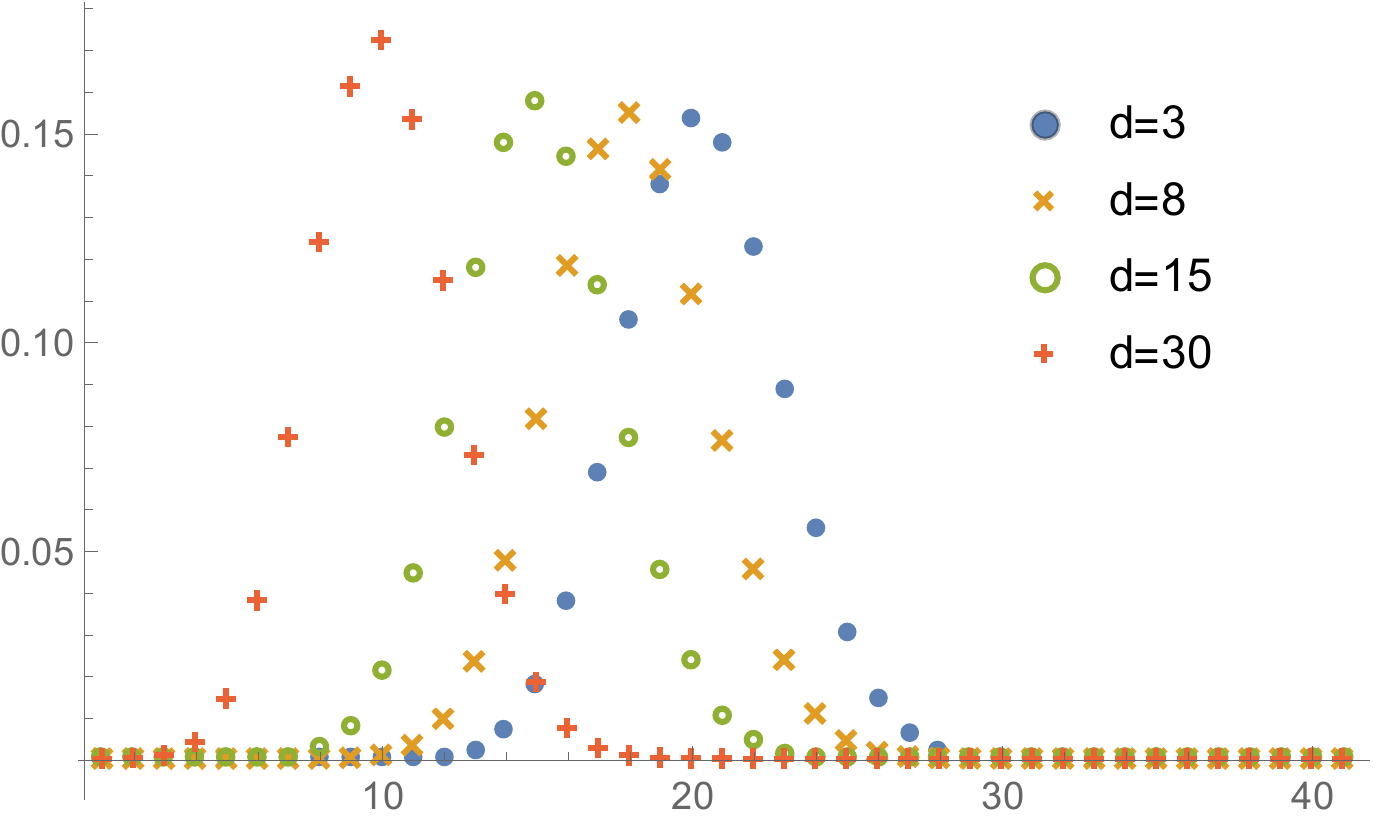}
	\end{subfigure}
	\caption{Plots of values $q_m^{d-1}(\tau)$ for $m=0,\ldots,40$ and parameters for the left plot: $d=3$ and times $\tau=0.1,0.2,0.3,0.5$ and for the right plot: $\tau=0.1$ and dimensions $d=3,8,15,30.$}
	\label{fig:coefq}
\end{figure}

We also need to analyse coefficients $q_m^{d-1}(\tau)$ for a fixed time $\tau$ and dimension $d$ and see which coefficients are relevant for the simulation (i.e. which coefficients are not negligible). To get a better understanding we therefore plot on Figure~\ref{fig:coefq} some of the coefficients for various times and dimensions.

Again we notice unimodality, which in this case  is not directly apparent from the  representation~\eqref{coef:q}. Additionally, when time $\tau$ increases, the distribution moves to the left, which is intuitively clear from a
description via death processes. Similarly, when the dimension $d$ increases, distribution moves to the left, but the effect is less profound.

Since we are limited with how small time $\tau$ we can take i.e. $\tau\ge0.1$, numerical calculations quickly show that the only relevant coefficients $q^{d-1}_m(\tau)$ are those with $m\le40$. All other coefficients will be essentially
equal to zero. Of course for larger values of $\tau$ and a larger dimension $d$ there might be even more essentially zero coefficients and there is no harm in setting them all equal to 0. Therefore, for a fixed time $\tau$ we will only
need to pre-calculate each coefficient $q^{d-1}_m(\tau)$ for $m\le40$. We do this calculations by taking sufficiently many terms in representation \eqref{coef:q} to get the desired accuracy. Additionally, we can stop calculations if we
reach a coefficient $q^{d-1}_m(\tau)$ which is already small enough and late enough so that the coefficients have already started decreasing. Again, we can set all succeeding coefficients to $0$. Once all relevant coefficients are
calculated, we can then use the standard inversion sampling to simulate a random variable $A^{d-1}_\infty(\tau)$ and use it in Algorithm~\ref{simSpB}.   

As we have seen, our algorithm does not work well for small values of $\tau$. The bottleneck of Algorithm~\ref{simSpB} is a simulation of $A_\infty^{d-1}(\tau)$. Hence, if we still want to use a variant of the algorithm, we need to resort
to some kind of an approximation and one option is using the following normal approximation for
$A_\infty^{d-1}(\tau)$, which was originally given in~\cite{Griff84}. Let $\beta=(d-2)t/2$ and $\eta=	\frac{\beta}{e^{\beta}-1}$. Then $A_\infty^{d-1}(\tau)$ is approximately distributed as a normal
$\operatorname{N}\left(\mu^{(\gamma,t)},\left(\sigma^{(\gamma,t)}\right)^2\right)$ random variable, where $\mu^{(\gamma,t)}=\frac{2\eta}{t}$ and $\left(\sigma^{(\gamma,t)}\right)^2=\frac{2\eta}{t}(\eta+\beta)^2\left(1 + \frac{\eta}{\eta+\beta}-2\eta \right)\beta^{-2}$. Alternatively, one can resort to using the approximation $\Theta\sim Q_{Approx}(\theta,\tau)$ and setting $X=\frac{1-\cos(\Theta)}{2}$ instead of steps \ref{stepAinf} and \ref{stepBeta}. Such an approximation actually seems to give better results than the normal approximation.

\section*{Acknowledgements}
AM and VM are supported by The Alan Turing Institute under the EPSRC grant EP/N510129/1;
AM supported by EPSRC grant EP/P003818/1 and the Turing Fellowship funded by the Programme
on Data-Centric Engineering of Lloyd’s Register Foundation; VM supported by the PhD scholarship of Department of Statistics, University of Warwick; GUB supported by CoNaCyT grant FC-2016-1946 and UNAM-DGAPA-PAPIIT grant IN114720.

\bibliographystyle{amsalpha}
\bibliography{source}

\providecommand{\bysame}{\leavevmode\hbox to3em{\hrulefill}\thinspace}
\providecommand{\MR}{\relax\ifhmode\unskip\space\fi MR }
\providecommand{\MRhref}[2]{%
  \href{http://www.ams.org/mathscinet-getitem?mr=#1}{#2}
}
\providecommand{\href}[2]{#2}
\begin{thebibliography}{KDPN00}

\bibitem[Bou16]{phylogeo}
R.~Bouckaert, \emph{Phylogeography by diffusion on a sphere: whole world
  phylogeography}, PeerJ \textbf{4} (2016), e2406.

\bibitem[BS98]{marinemigration}
D.R. Brillinger and B.S. Stewart, \emph{Elephant-{S}eal {M}ovements:
  {M}odelling {M}igration}, The Canadian Journal of Statistics / La Revue
  Canadienne de Statistique \textbf{26} (1998), no.~3, 431--443.

\bibitem[Cai04]{Cail2004}
J.-M. Caillol, \emph{Random walks on hyperspheres of arbitrary dimensions},
  Journal of Physics A: Mathematical and General \textbf{37} (2004), no.~9,
  3077.

\bibitem[CEE10]{sim2dimsphere}
T.~Carlsson, T.~Ekholm, and C.~Elvingson, \emph{Algorithm for generating a
  {B}rownian motion on a sphere}, Journal of Physics A: Mathematical and
  Theoretical \textbf{43} (2010), no.~50, 505001.

\bibitem[Far02]{fara}
J.~Faraudo, \emph{Diffusion equation on curved surfaces. i. {T}heory and
  application to biological membranes}, The Journal of Chemical Physics
  \textbf{116} (2002), no.~13, 5831--5841.

\bibitem[GL83]{GriffLi83}
R.~C. Griffiths and W.-H. Li, \emph{Simulating allele frequencies in a
  population and the genetic differentiation of populations under mutation
  pressure}, Theoretical Population Biology \textbf{23} (1983), no.~1, 19 --
  33.

\bibitem[Gri84]{Griff84}
R.~C. Griffiths, \emph{Asymptotic line-of-descent distributions}, Journal of
  Mathematical Biology \textbf{21} (1984), no.~1, 67--75.

\bibitem[GS10]{GriffSpanoDiffCoal}
R.~C. {Griffiths} and D.~Spanò, \emph{{Diffusion processes and coalescent
  trees}}, ArXiv e-prints (2010).

\bibitem[GSS12]{gaussForSph}
A.~Ghosh, J.~Samuel, and S.~Sinha, \emph{A {"Gaussian"} for diffusion on the
  sphere}, EPL (Europhysics Letters) \textbf{98} (2012), no.~3, 30003.

\bibitem[Hsu02]{hsumanifolds}
E.P. Hsu, \emph{Stochastic analysis on manifolds}, Graduate Studies in
  Mathematics 38, vol.~38, American Mathematical Society, 2002.

\bibitem[JS17]{exsimWF}
P.A. Jenkins and D.~Span\`{o}, \emph{Exact simulation of the
  {W}right–{F}isher diffusion}, Ann. Appl. Probab. \textbf{27} (2017), no.~3,
  1478--1509.

\bibitem[KDPN00]{KrishnaFluor}
M.~M.~G. Krishna, Ranjan Das, N.~Periasamy, and Rajaram Nityananda,
  \emph{Translational diffusion of fluorescent probes on a sphere: {M}onte
  {C}arlo simulations, theory, and fluorescence anisotropy experiment}, The
  Journal of Chemical Physics \textbf{112} (2000), no.~19, 8502--8514.

\bibitem[KT81]{KarTay}
S.~Karlin and H.~M. Taylor, \emph{A second course in stochastic processes},
  Academic Press, 1981.

\bibitem[LTT08]{BactLi1}
G.~Li, L.-K. Tam, and J.~X. Tang, \emph{Amplified effect of {B}rownian motion
  in bacterial near-surface swimming}, Proceedings of the National Academy of
  Sciences \textbf{105} (2008), no.~47, 18355--18359.

\bibitem[MMU20]{simViaWF}
A.~{Mijatovi{\'c}}, V.~{Mramor}, and G.~{Uribe Bravo}, \emph{{A note on the
  exact simulation of spherical {B}rownian motion}}, Statistics \& Probability
  Letters \textbf{165} (2020), 108836.

\bibitem[NEE03]{sim4dimsphere}
J.~Nissfolk, T.~Ekholm, and C.~Elvingson, \emph{Brownian dynamics simulations
  on a hypersphere in 4-space}, The Journal of Chemical Physics \textbf{119}
  (2003), no.~13, 6423--6432.

\end{thebibliography}
     
\end{document}